\documentclass[aps,prb,letterpaper,amsmath,amssymb,reprint,superscriptaddress]{revtex4-1}
\usepackage{graphicx}

\begin{document}
\title{Itinerant and local magnetic moments in ferrimagnetic Mn$_2$CoGa thin films probed by x-ray magnetic linear dichroism: experiment and \textit{ab initio} theory}
\author{Markus Meinert}
\email{meinert@physik.uni-bielefeld.de}
\author{Jan-Michael Schmalhorst}
\author{Christoph Klewe}
\author{G\"unter Reiss}
\affiliation{Thin Films and Physics of Nanostructures, Department of Physics, Bielefeld University, D-33501 Bielefeld, Germany}
\author{Elke Arenholz}
\affiliation{Advanced Light Source, Lawrence Berkeley National Laboratory, CA 94720, USA}
\author{Tim B\"ohnert}
\author{Kornelius Nielsch}
\affiliation{Institute of Applied Physics, University of Hamburg, Jungiusstrasse 11, D-20355 Hamburg, Germany}

\date{\today}

\begin{abstract}
Epitaxial thin films of the half-metallic X$_a$-compound Mn$_2$CoGa (Hg$_2$CuTi prototype) were prepared by dc magnetron co-sputtering with different heat treatments on MgO (001) substrates. High-quality films with a bulk magnetization of $1.95(5)\,\mu_\text{B}$ per unit cell were obtained. The L$_{3,2}$ x-ray magnetic circular dichroism spectra agree with calculations based on density functional theory (DFT) and reveal the antiparallel alignment of the two inequivalent Mn moments. X-ray magnetic linear dichroism, in good agreement with theory as well, allows to distinguish between itinerant and local Mn moments. Based on non-collinear spin DFT it is shown that one of the two Mn moments has local character, whereas the other Mn moment and the Co moment are itinerant.
\end{abstract}

\maketitle

The recently discovered class of Mn$_2$\textit{YZ} compounds that crystallize in the X$_a$ (also called inverse Heusler, prototype Hg$_2$CuTi) structure has been the subject of considerable theoretical and experimental activities with \textit{Y} = Fe, Co, Ni, Cu and \textit{Z} a group III, IV, or V element. \cite{Luo08_1, Liu08, Xing08, Lakshmi05, Dai06, Helmholdt87, Luo09_1, Luo09_2, Wei10, Li09, Winterlik11} It can be imagined as four interpenetrating fcc lattices with the basis vectors A = (0, 0, 0), B = (1/4, 1/4, 1/4), C = (1/2, 1/2, 1/2), and D = (3/4, 3/4, 3/4), where Mn occupies the non-equivalent B and C sites as nearest neighbors. In our nomenclature \textit{Y} resides on the A site and \textit{Z} is on the D site. In the case of Mn$_2$CoGa (and its isoelectronic relatives Mn$_2$CoAl and Mn$_2$CoIn) the total magnetic moment is $2\,\mu_\text{B}$/f.u., in agreement with the Slater-Pauling rule for half-metallic Heusler compounds. \cite{Liu08, Galanakis02} These compounds are governed by a strong direct exchange interaction between the antiparallel Mn moments, which results in Curie temperatures of possibly more than 800\,K. \cite{Meinert11_1} Ferrimagnetic (inverse) Heusler compounds are promising parent materials for doping, aiming at further reduction of the magnetic moment. \cite{Meinert11_2, Klaer11}

To date, the inverse Heusler compounds were studied only in the bulk. For many practical applications, such as in tunnel or giant magnetoresistance (TMR, GMR) devices, thin films are necessary. In this article we report on the first preparation and characterization of thin films from the class of the inverse Heusler compounds.

We prepared epitaxial thin films of Mn$_2$CoGa with (001) orientation by dc magnetron co-sputtering on MgO (001) substrates. A Mn$_{50}$Ga$_{50}$ target and an elemental Co target were used for the deposition. The resulting Mn:Ga ratio in the films was 2.2:1, as determined by x-ray fluorescence. Co was added to match the Ga content, i.e., the stoichiometry of the unit cell can be written as Mn$_{2.1}$Co$_{0.95}$Ga$_{0.95}$.

Among the various heat treatments tested, deposition at 200$^\circ$C and \textit{in situ} post-annealing at 550$^\circ$C was found to provide optimal film quality. The lattice parameter perpendicular to the surface was 5.81\,\AA{}, which is slighty smaller than the bulk value of 5.86\,\AA{}.\cite{Liu08} A small tetragonal distortion of the film is induced by the lattice mismatch with the substrate, hence the lattice is expanded in the film plane and compressed perpendicular to the plane. The bulk magnetization measured by a superconducting quantum interference device (SQUID) corresponds to $1.95(5)\,\mu_B$ / unit cell, which is very close to the bulk value. No significant change of the magnetization between 5\,K and room temperature was observed, which is consistent with a Curie temperature higher than 600\,K. 

X-ray absorption (XAS) measurements were performed at BL4.0.2 of the Advanced Light Source in Berkeley, CA, USA. X-ray magnetic circular (XMCD) and linear dichroism (XMLD) measurements were taken at room temperature in x-ray transmission through the film by collecting the visible and ultraviolet light fluorescence from the substrate with a photodiode. \cite{Kallmayer07} The sample was saturated with a magnetic field of 0.6\,T and the circular or linear polarization degree was 90\,\% and 100\,\%, respectively.

We computed the XAS, XMCD and XMLD using density functional theory. The full potential linearized augmented plane waves (FLAPW) method, implemented in the Elk code,\cite{elk} was used. The experimental bulk lattice parameter was chosen for the calculations; the small distortion and off-stoichiometry have negligible influence. The Brillouin zone integration was performed on a $16 \times 16 \times 16$ k-point mesh in the irreducible wedge, the Perdew-Burke-Ernzerhof functional\cite{PBE} was chosen for exchange and correlation and spin-orbit coupling was included in a second-variational scheme. The absorption and dichroic spectra were calculated within a first order optical response formalism, i.e., core-hole correlations were not taken into account. A half-metallic ground-state was obtained with a total spin magnetic moment of 2\,$\mu_\text{B}$/f.u., and site resolved spin (orbital) moments as follows: Co 1.03\,$\mu_\text{B}$ (0.046\,$\mu_\text{B}$), Mn(B) 2.91\,$\mu_\text{B}$ (0.011\,$\mu_\text{B}$), and Mn(C) $-1.93$\,$\mu_\text{B}$ ($-0.019$\,$\mu_\text{B}$). A detailed discussion of the electronic structure is given in Ref. \onlinecite{Liu08}.

\begin{figure}[t]
\includegraphics[width=9cm]{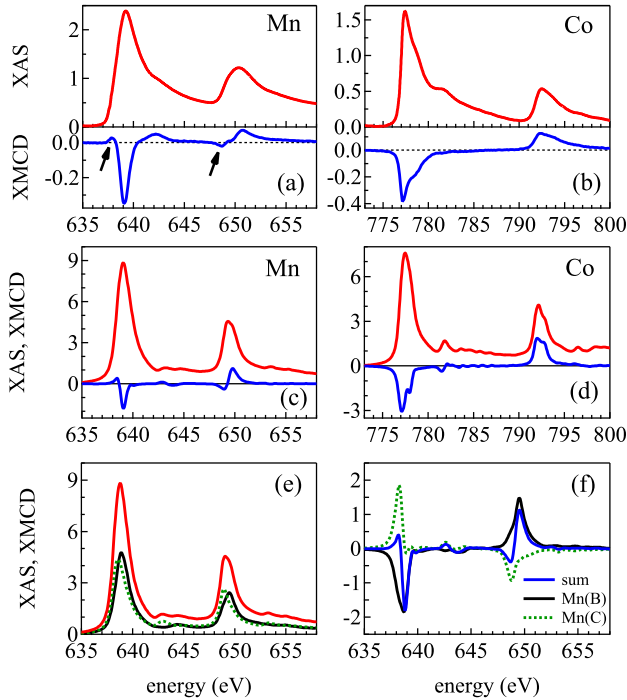}
\caption{Top: experimental XAS and XMCD spectra of (a): Mn and (b): Co in Mn$_2$CoGa. Middle: theoretical XAS and XMCD spectra of Mn$_2$CoGa. (c): Mn XAS and XMCD. (d): Co XAS and XMCD. Bottom: decomposition of the Mn XAS (e) and XMCD (f) for the two inequivalent Mn sites. The theoretical spectra are normalized to 1 about 40\,eV above the L$_3$ edge and are shifted to match the experimental absorption onset at L$_3$.}
\label{Fig1}
\end{figure}

The experimental x-ray absorption and circular dichroism spectra are shown in Fig. \ref{Fig1} (a) and (b). Both x-ray absorption spectra have the typical shape of a metallic system without pronounced multiplets. However, the XMCD spectrum of Mn shows some uncommon features (see arrows in Fig. \ref{Fig1}a). The Co XAS exhibits fine structures at the L$_3$ and L$_2$ resonances. There is a weak shoulder about 2.6\,eV above threshold and a more pronounced one at 5\,eV above threshold. The Co XMCD spectrum reflects the shoulder in the XAS. The Co and (effective) Mn moments are parallel. All these features are reproduced by the \textit{ab initio} calculations (Fig. \ref{Fig1} (c) and (d)), which are broadened with a Lorentzian of 0.3\,eV width to account for lifetime effects. We can thus identify the features in the spectra as band structure effects. The 5\,eV feature in the Co XAS results from transitions into an \textit{s-d} hybridized state of Co and Ga. It is commonly observed for Co in Co$_2$\textit{YZ} type Heusler compounds, but its position depends on the \textit{Z} element. The asymmetric line shape and the broad tails of the resonances are a consequence of 2\textit{p}-3\textit{d} \textit{e-e} correlation,\cite{Pardini11} which is neglected in our simulations. Electron-hole correlations can significantly alter the shape of the XAS or XMCD spectra of 3\textit{d} transition elements, even in a metallic environment. \cite{Meinert11_3} Thus, the good agreement of our calculations with the experimental spectra indicates an effective screening of the 2\textit{p} core-hole.

In Fig. \ref{Fig1} (e) and (f) we show the decomposition of the calculated XAS and XMCD into the Mn(B) and Mn(C) components. We find that the core levels of Mn(B) and Mn(C) are slightly shifted (about 0.15\,eV) against each other. The shapes of the spectra as well as the branching ratios are different, the Mn(B) branching ratio is significantly larger than the one of Mn(C). The decomposition of the XMCD spectrum shows two different signals with opposite signs. The antiparallel Mn(C) contribution is responsible for the features marked in the experimental spectrum. These features are less pronounced in the experimental spectrum, which indicates a smaller core-level shift than the calculated one.

A sum rule analysis was performed to obtain the spin and orbital magnetic moments from the XMCD data. Details of the procedure are given in Ref. \onlinecite{Schmalhorst09}. The resulting magnetic moment ratios are: $m_\text{spin}^\text{Mn}$ / $m_\text{spin}^\text{Co} = 0.48$, $m_\text{orb}^\text{Mn}$ / $m_\text{spin}^\text{Mn} = -0.013$, $m_\text{orb}^\text{Co}$ / $m_\text{spin}^\text{Co} = 0.055$. Using the bulk magnetization we derive the element specific moments. The average Mn spin moment is $0.47\,\mu_\textit{B}$ per atom and the Co spin moment is $0.98\,\mu_\textit{B}$ per atom. The average orbital moment of Mn is -0.006\,$\mu_\text{B}$ per atom, being antiparallel to the spin magnetic moment. For Co we find 0.055\,$\mu_\text{B}$ for the orbital moment. In this analysis the apparent Mn spin moment has been multiplied by 1.5 to compensate the 2\textit{p}$_{1/2}$ - 2\textit{p}$_{3/2}$ channel mixing, as suggested by D\"urr \textit{et al}. \cite{Duerr97} These values match the theoretical values within the errors. Both the positive Co orbital moment as well as the small negative Mn orbital moment are in agreement with the calculation. The orbital moments of all atoms are parallel to the respective spin moments, but the orbital moment of Mn(C) is larger than the one of Mn(B), resulting in the effectively antiparallel alignment.

The single crystalline character of epitaxial films allows to make use of the anisotropic x-ray magnetic linear dichroism. \cite{Kunes03} In general, the XMLD is proportional to the spin moment squared. \cite{Freeman06, Arenholz10} In contrast to XMCD, XMLD is only sensitive to the direction of the spin moments, not their orientation. This allows to probe antiferromagnetic and ferrimagnetic materials with XMLD. For local moments, the core-level exchange splitting is stronger than for itinerant moments, which leads to an enhanced XMLD amplitude. Therefore, XMLD can be used as a probe for the locality of magnetic moments by comparison with reference systems. Because XMLD is essentially given as the difference between $\Delta m = 0$ and the averaged $\Delta m = \pm 1$ transitions, it is a sensitive probe of the local crystal field.

It was shown that the Mn moment has a local character in the Heusler compounds Co$_2$MnSi (CMS) and Co$_2$MnAl (CMA).\cite{Telling08} K\"ubler \textit{et al.} proposed an exclusion of minority \textit{d} electrons from the environment of Mn, giving rise to a local moment composed of itinerant electrons.\cite{Kuebler83} A similar mechanism can give rise to a local Mn(B) moment in Mn$_2$CoGa.\cite{Liu08} Therefore, we chose CMS as a reference system with similar crystal structure for local moments. Mn$_2$VGa (MVG), also crystallizing in the Heusler structure, is postulated to be itinerant, and is chosen as a reference system for itinerant Mn moments.

\begin{figure}[b]
\includegraphics[width=8.6cm]{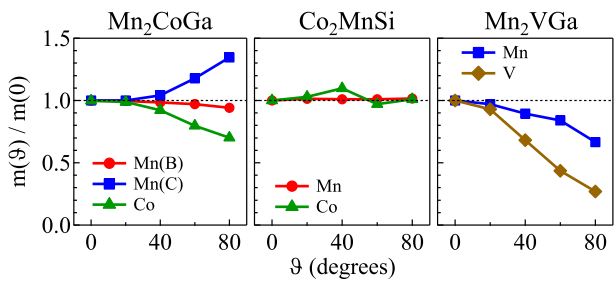}
\caption{Calculated relative change of the magnetic moments for non-collinear configurations. The spin moment under investigation is tilted out of the common axis by $\vartheta$.}
\label{Fig2}
\end{figure}

A simple theoretical test for the (non-)locality of spin moments is based on non-collinear spin configurations. We performed self-consistent calculations for non-collinear configurations (without spin-orbit coupling) in which the magnetic moment of interest was tilted by an angle $\vartheta$ out of the common magnetization axis. Only the directions were fixed, and the magnitudes were determined self-consistently. A local moment would not change in magnitude when tilted. In Fig. \ref{Fig2} the relative changes of the magnetic moments for Mn$_2$CoGa and the reference systems CMS and MVG are shown. In Mn$_2$CoGa, Mn(B) has a weak dependence on $\vartheta$, whereas Mn(C) and Co change significantly on tilting: Mn(B) has local character, whereas Mn(C) and Co are rather itinerant. Both the Co and the Mn moment in CMS have weak or no dependence on the tilt angle, showing clearly the locality of both moments. MVG in contrast, is an itinerant system; both the Mn and the V moment depend strongly on $\vartheta$. Mn$_2$CoGa has a more complex magnetic structure than the reference compounds, being a hybrid between itinerant and local magnetism. Local moment systems can be described within the Heisenberg model. This has been successfully applied to explain the Curie temperatures in CMS and related compounds.\cite{Thoene09} For MVG, this model underestimates the Curie temperature,\cite{Meinert11_1} similar to fcc Ni.\cite{Liechtenstein87} This can be seen as experimental evidence for the itinerancy of MVG. Consequently, we expect significant deviation of experimental Curie temperatures from theoretical values for Mn$_2$CoGa.

\begin{figure}[t]
\includegraphics[width=8.6cm]{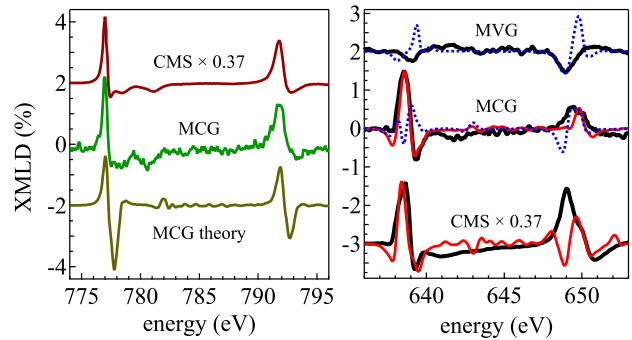}
\caption{Left: Experimental and theoretical Co XMLD spectra of Mn$_2$CoGa and Co$_2$MnSi (experimental spectrum from Ref. \onlinecite{Telling08}). Right: Experimental (black solid lines) and theoretical (thin lines) Mn XMLD spectra of Mn$_2$VGa, Mn$_2$CoGa and Co$_2$MnSi (experimental spectrum from Ref. \onlinecite{Telling08}). Mn(B) type spectra are solid red, Mn(C) type spectra are dotted blue. The XMLD is normalized to the L$_3$ resonance height. All theoretical spectra are shifted and expanded to match the experimental absorption onset at L$_3$ and the L$_{3,2}$ spin-orbit splitting. They are scaled to match the experimental intensities.}
\label{Fig3}
\end{figure}

We have performed XMLD measurements for Co and Mn along the [110] direction of the film. In Fig. \ref{Fig3} we show the experimental and theoretical spectra of Mn$_2$CoGa and the reference compounds. All XMLD data were taken at the same beamline and are directly comparable in terms of energy resolution.

The Co XMLD of Mn$_2$CoGa is very similar in shape to the signal of CMS, all fine details are reproduced. The computed spectrum of Mn$_2$CoGa resembles the general shape of the experimental data, although the negative contributions are overestimated. These are in the tails of the resonances, in which \textit{e}-\textit{e} correlation plays a role, which we neglect as stated above. The local crystal fields are consequently similar in Mn$_2$CoGa and CMS, and the \textit{ab initio} calculation is able to describe these reasonably well.

For Mn, we find that the Mn$_2$CoGa and the CMS signals are virtually equal at L$_3$. At L$_2$ however, they are somewhat different. Mn$_2$CoGa has an overall less pronounced structure and less intensity here. The MVG signal is much weaker and has an entirely different shape, which indicates different crystal fields acting on Mn on a B or C position. The computed spectra of Mn(B) in Mn$_2$CoGa and for CMS resemble the experimental data at L$_3$ very well. At L$_2$, significant deviation is observed, particularly for CMS. The main peak at L$_2$ in CMS stems from a feature in the XAS that was assigned to an atomic multiplet, that survives the band formation and corroborates the locality of the moment.\cite{Telling08} In Mn$_2$CoGa this feature is less pronounced, leading to a better agreement of experiment and theory. Less locality of the Mn(B) moment in comparison to CMS can be inferred from that. The influence of the Mn(C) spectrum in Mn$_2$CoGa can not be traced in the experimental data. The calculated Mn(C) spectrum is, however, very similar to the computed XMLD of MVG. This, in turn, agrees only modestly with experiment. Because of the similarity of the computed spectra, we assume that the actual Mn(C) contribution would have similar shape as the measured MVG spectrum. The Mn$_2$CoGa XMLD is, in conclusion, clearly dominated by the Mn(B) signal.

\begin{figure}[t]
\includegraphics[width=8.6cm]{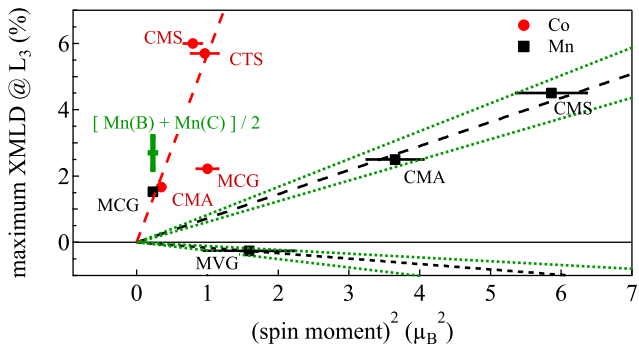}
\caption{XMLD vs. $m_s^2$ for various Mn and Co containing (inverse) Heusler compounds: Co$_2$MnSi (CMS), Co$_2$MnAl (CMA), Co$_2$TiSn (CTS), Mn$_2$VGa (MVG), and Mn$_2$CoGa (MCG).}
\label{Fig4}
\end{figure}

Now we turn to the observed intensities of the XMLD signals. Fig. \ref{Fig4} shows a comparison of the maximum XMLD signals (defined as $(I^{||} - I^{\perp})|_\text{max} / [(I^{||} + I^{\perp})/2]|_\text{max}$) at the L$_3$ edges versus the squared spin magnetic moments of Co and Mn for CMS, CMA, Co$_2$TiSn (CTS), MVG, and Mn$_2$CoGa. The CTS data were taken from Ref. \onlinecite{Meinert11_3}. The Co XMLD amplitudes are close to a common line for CMS, CMA, and CTS. CMS is a bit above though, indicating a stronger locality of the Co moment in CMS than in CMA or CTS. The Mn$_2$CoGa signal is about a factor of 2.5 smaller than expected from the references. In agreement with the locality test described above, this shows the itinerancy of the Co moment in Mn$_2$CoGa. Because of the antiparallel Mn moments, the Mn XMLD of Mn$_2$CoGa is very strong compared to the Mn spin moment, and it is far off the line given by CMS and CMA.

With the linear fits through the CMA and CMS points as a guide for local Mn moments and through the MVG point for an itinerant system we can predict the Mn XMLD amplitude of Mn$_2$CoGa. We treat the Mn XMLD of Mn$_2$CoGa as a superposition of the spectra from CMA/CMS and MVG. Our FLAPW calculation gives a Mn(B)/Mn(C) spin moment ratio of $-1.5$. With this value and the measured sum $m^\text{Mn(B)}_\text{s} + m^\text{Mn(C)}_\text{s} \approx 0.94\,\mu_\text{B}$ we obtain $m^\text{Mn(B)}_\text{s} = 2.82\,\mu_\text{B}$  and $m^\text{Mn(C)}_\text{s} = -1.88\,\mu_\text{B}$. According to the errors of the magnetic moments of the reference data, we expect an XMLD of $(2.7 \pm 0.5)\%$ for Mn$_2$CoGa. The measured value of 1.53\% is clearly below this range; the ratio determined directly from the XMLD is $-1.7$, which leads to $m^\text{Mn(B)}_\text{s} = 2.28\,\mu_\text{B}$  and $m^\text{Mn(C)}_\text{s} = -1.34\,\mu_\text{B}$. Though this is still reasonable, it seems much more likely that the lower XMLD in Mn$_2$CoGa indicates a lower degree of Mn(B) spin moment locality than in CMS. However, the Mn(B) moment is clearly not purely itinerant.

In summary, we have prepared epitaxial films of the ferrimagnetic inverse Heusler compound Mn$_2$CoGa by co-sputtering and obtained good film quality by deposition at 200$^\circ$C and \textit{in situ} post-annealing at 550$^\circ$C. We found good agreement of the experimental L$_{3,2}$ x-ray absorption and dichroism spectra with \textit{ab initio} calculations within independent particle theory. The total and element resolved magnetic moments are close to theoretical values. X-ray magnetic linear dichroism spectra were taken to provide information on the locality of the Co and Mn moments. Non-collinear electronic structure calculations provided the footing for the interpretation of the observed XMLD amplitudes. The locality of the Mn(B) moment is not as pronounced as in Co$_2$MnSi, the Co and Mn(C) moments have clearly itinerant character. Density functional theory has proven to be an indispensable tool for the interpretation of complex spectroscopic data.

The authors gratefully acknowledge financial support from Bundesministerium f\"ur Bildung und Forschung (BMBF) and Deutsche Forschungsgemeinschaft (DFG). They thank for the opportunity to work at the Advanced Light Source, Berkeley, CA, USA, which is supported by the Director, Office of Science, Office of Basic Energy Sciences, of the US Department of Energy under Contract No DE-AC02-05CH11231. K.N. and T.B. gratefully acknowledge the financial support by the state of Hamburg via the cluster of excellence LEXI Nanospintronics. Special thanks go to the developers of the Elk code.


\begin{thebibliography}{50}
\bibitem{Luo08_1} H. Z. Luo \textit{et al.}, J. Appl. Phys. {\bf 103}, 083908 (2008).
\bibitem{Liu08} G. D. Liu \textit{et al.}, Phys. Rev. B {\bf 77}, 014424 (2008).
\bibitem{Xing08} N. Xing, H. Li, J. Dong, R. Long, and C. Zhang, Computational Materials Science {\bf 42}, 600 (2008).
\bibitem{Lakshmi05} N. Lakshmi, R. K. Sharma, and K. Venugopalan, Hyperfine Interactions {\bf 160}, 227 (2005).
\bibitem{Dai06} X. Dai, G. Liu, L. Chen, J. Chen, and G. Wu, Solid State Comm. {\bf 140}, 533 (2006).
\bibitem{Helmholdt87} R. B. Helmholdt and K. H. J. Buschow, J. Less-Common Met. {\bf 128}, 167 (1987).
\bibitem{Luo09_1} H. Luo \textit{et al.}, J. Magn. Magn. Mater. {\bf 321}, 4063 (2009).
\bibitem{Luo09_2} H. Luo \textit{et al.}, J. Appl. Phys. {\bf 105}, 103903 (2009).
\bibitem{Wei10} X. P. Wei \textit{et al.}, J. Magn. Magn. Mater. {\bf 322}, 3204 (2010).
\bibitem{Li09} S. T. Li, Z. Ren, X. H. Zhang, and C. M. Cao, Physica B: Cond. Matter {\bf 404}, 1965 (2009).
\bibitem{Winterlik11} J. Winterlik \textit{et al.}, Phys. Rev. B {\bf 83}, 174448 (2011).
\bibitem{Galanakis02} I. Galanakis, P. H. Dederichs, and N. Papanikolaou, Phys. Rev. B {\bf 66}, 174429 (2002).
\bibitem{Meinert11_1} M. Meinert, J. M. Schmalhorst, and G. Reiss, J. Phys.: Condens. Matt. {\bf 23}, 116005 (2011).
\bibitem{Meinert11_2} M. Meinert, J. M. Schmalhorst, G. Reiss, and E. Arenholz, J. Phys. D: Appl. Phys. \textbf{44}, 215003 (2011).
\bibitem{Klaer11} P. Klaer \textit{et al.}, Appl. Phys. Lett. {\bf 98}, 212510 (2011).
\bibitem{Kallmayer07} M. Kallmayer \textit{et al.}, J. Phys. D: Appl. Phys. {\bf 40} 1552 (2007).
\bibitem{elk} Elk version 1.3.24, http://elk.sourceforge.net.
\bibitem{PBE} J. P. Perdew, K. Burke, M. Ernzerhof, Phys. Rev. Lett. {\bf 77}, 3865 (1996).
\bibitem{Pardini11} L. Pardini, V. Bellini, and F. Manghi, J. Phys.: Condens. Matter \textbf{23}, 215601 (2011).
\bibitem{Meinert11_3} M. Meinert \textit{et al.}, Phys. Rev. B \textbf{83}, 064412 (2011).
\bibitem{Schmalhorst09} J. Schmalhorst \textit{et al.}, J. Appl. Phys. {\bf 105}, 053906 (2009).
\bibitem{Duerr97} H. A. D\"urr \textit{et al.}, Phys. Rev. B \textbf{56}, 8156 (1997).
\bibitem{Kunes03} J. Kunes and P. M. Oppeneer, Phys. Rev. B {\bf 67}, 024431 (2003).
\bibitem{Freeman06} A. A. Freeman \textit{et al.}, Phys. Rev. B {\bf 73}, 233303 (2006).
\bibitem{Arenholz10} E. Arenholz, G. van der Laan, A. McClure, and Y. Idzerda, Phys. Rev. B {\bf 82}, 180405(R) (2010).
\bibitem{Telling08} N. D. Telling \textit{et al.}, Phys. Rev. B {\bf 78}, 184438 (2008); the corresponding spin magnetic moments are $m^\text{CMS}_\text{Co} = 0.89\,\mu_\text{B}$, $m^\text{CMS}_\text{Mn} = 2.42\,\mu_\text{B}$, $m^\text{CMA}_\text{Co} = 0.59\,\mu_\text{B}$, $m^\text{CMA}_\text{Mn} = 1.91\,\mu_\text{B}$. Errors of $\pm10\%$ were assumed for these values. N. D. Telling, private communication.
\bibitem{Kuebler83} J. K\"ubler, A. R. Williams, and C. B. Sommers, Phys. Rev. B {\bf 28}, 1745 (1983).
\bibitem{Thoene09} J. Thoene \textit{et al.}, J. Phys. D: Appl. Phys. {\bf 42}, 084013 (2009).
\bibitem{Liechtenstein87} A. I. Liechtenstein, M.I. Katsnelson, V. P. Antropov, and V. A. Gubanov, J. Magn. Magn. Mat. {\bf 67}, 65 (1987).
\end{thebibliography}
\end{document}